\documentclass[prd,twocolumn,nofootinbib]{revtex4}
\usepackage{graphicx, epsfig}
\usepackage{color}
\usepackage{mathrsfs}
\usepackage{bm}
\usepackage{amsmath,amssymb}
\usepackage{mathrsfs}
\usepackage[caption=false]{subfig}
\usepackage{hyperref}
\usepackage[normalem]{ulem}
\usepackage{mathtools}
\usepackage[x11names]{xcolor}

\definecolor{fashionfuchsia}{rgb}{0.96, 0.0, 0.63}
\colorlet{no_so_fashion_purple}{blue!50!red}

\newcommand{\be}{\begin{equation}}
\newcommand{\ee}{\end{equation}}
\newcommand{\ba}{\begin{eqnarray}}
\newcommand{\ea}{\end{eqnarray}}

\newcommand{\nn}{\nonumber}
\newcommand{\half}{\frac{1}{2}}
\newcommand{\fourth}{\frac{1}{4}}

\newcommand{\hatna}{{\hat n}^a}

\newcommand{\Lag}{{\cal L}}

\newcommand{\calE}{{\cal E}}

\def\half{\frac{1}{2}}

\begin{document}
\title{Construction of non-Abelian electric strings}
\author{Tanmay Vachaspati}
\affiliation{
$^*$Physics Department, Arizona State University, Tempe,  Arizona 85287, USA.
}

\begin{abstract}
We detail the construction of electric string solutions in $SU(2)$ Yang-Mills-Higgs theory 
with a scalar in the fundamental representation and discuss the properties of the solution.
We show that Schwinger gluon pair production in the electric string background is absent. 
A similar construction in other models, such as with an adjoint scalar field and the electroweak
model, does not yield solutions.
\end{abstract}

\maketitle

\section{Introduction}
\label{intro}

A homogeneous electric field in Maxwell electrodynamics corresponds to the familiar
gauge potential
\be
A^\mu = (- E z ,0,0,0).
\label{maxwellA}
\ee
where $E$ is the electric field strength. When coupled to external charges, the electric field 
is known to decay by Schwinger pair production~\cite{Schwinger:1951nm}. Similarly, an $SU(2)$ non-Abelian
electric field\footnote{We will explicitly only consider $SU(2)$ non-Abelian gauge theory. The
solutions may be embedded in theories with larger gauge groups.},
can be derived from the gauge potential
\be
A^{\mu a}= (- E z ,0,0,0) \delta^{a3}
\label{nAmaxwellA}
\ee
where $a=1,2,3$ is the group index. Unlike in the Maxwell case, it is not necessary to introduce 
``external'' charges as even the pure non-Abelian gauge theory includes charged quanta
(``gluons''). Schwinger pair production of gluons will cause the non-Abelian electric
field to decay rapidly~\cite{Matinyan:1976mp,Brown:1979bv,Yildiz:1979vv,Ambjorn:1981qc,
Ambjorn:1982nd,Nayak:2005yv,Cooper:2005rk,Cooper:2008vy,Nair:2010ea,
Kim:2011jw,Ilderton:2021zej,Huet:2014mta,Ragsdale:2017wgi,Karabali:2019ucc,
Cardona:2021ovn}.
However the story for non-Abelian electric fields is more subtle, as 
embedding the Maxwell gauge potential into the non-Abelian theory is not the only way
 to obtain a non-Abelian electric field. As shown by
Brown and Weisberger (BW)~\cite{Brown:1979bv}, and described in Sec.~\ref{gaugefields}, 
there is a one parameter set of gauge inequivalent gauge fields that all lead to the 
same homogeneous electric field. An analysis of Schwinger pair production in such 
gauge field backgrounds shows that Schwinger gluon production is absent~\cite{Vachaspati:2022ktr}. 

An issue with BW gauge fields is that, unlike Eq.~\ref{nAmaxwellA}, they are not classical 
solutions of the vacuum equations of non-Abelian gauge theory; instead they require 
sources. A possibility is that quantum backreaction on the classical dynamics effectively
provides such sources but this is difficult to show. A second possibility, one that has 
been reported in Ref.~\cite{Vachaspati:2022gco} and that we will detail in this paper, is that suitable 
sources can be provided by an external classical field, such as a scalar field. Then
the BW gauge fields are solutions of the Yang-Mills-Higgs classical equations of
motion, much like other classical solutions such as strings and magnetic monopoles
\cite{Vilenkin:2000jqa}, but a key difference is that the solution contains a flux of electric field
instead of a magnetic field. Such solutions are called ``electric strings''.

The electric string solution presented in Ref.~\cite{Vachaspati:2022gco} was arrived at by using a
certain amount of guesswork. In the present paper, we present some rationale
for the guesses, in addition to exploring certain other issues.
The starting point for our discussions is to consider a homogeneous BW electric field,
discussed in Sec.~\ref{gaugefields}. In Sec.~\ref{homoE} we show that a homogeneous 
BW electric field can indeed be sourced by a scalar field that transforms in the fundamental 
representation of $SU(2)$. The necessary scalar field configuration also solves its own
equation of motion but only for certain parameters. 

Encouraged by the case of the homogeneous electric field, we turn our
attention to ``electric string'' solutions in which the electric field is localized
to a tubular region. In Sec.~\ref{estring} we construct such a solution. 
We find that the tubular electric field is wrapped by magnetic fields in
an oscillating pattern with a slow asymptotic fall off.

The question of Schwinger gluon production in the background of an electric
string is considered in Sec.~\ref{schwinger} and, as in the homogeneous
electric field case, this process is absent. A classical stability analysis of the
electric string solution is postponed for future work.

The existence of an electric string solution in a Yang-Mills with a fundamental Higgs
naturally raises the question if such strings can arise in other theories. In
Appendix~\ref{adjoint} we examine the case when the scalar field is in the
adjoint representation of $SU(2)$ and show that the solution does not exist.
Similarly in Appendix~\ref{electroweak} we examine if the solution can be
found in the electroweak model that has an additional $U(1)$ hypercharge gauge 
field, and there too we find that a solution does not exist. These no-go
results though are based on certain assumptions about the structure of
the solutions and it is possible that a more general analysis might successfully
find solutions.


Our conclusions are summarized in Sec.~\ref{conclusions} and some helpful 
formulae are listed in Appendix~\ref{someformulae}.

\section{Gauge fields}
\label{gaugefields}

\subsection{BW homogeneous gauge fields}

The non-vanishing $SU(2)$ gauge fields that give a
homogeneous electric field are~\cite{Brown:1979bv},
\be
W_\mu^1 = \frac{\Omega}{g} \partial_\mu t, \ \ 
W_\mu^2 = -\frac{E}{\Omega} \partial_\mu z, \ \ 
W_\mu^3 = 0.
\label{BWfields}
\ee
where $\Omega$ is a parameter. We will assume without loss of
generality that $\Omega > 0$.
The field strength is found from
\be
W_{\mu\nu}^a = \partial_\mu W_\nu^a - \partial_\nu W_\mu^a 
+ g \epsilon^{abc} W_\mu^b W_\nu^c
\label{fsdefn}
\ee
and the non-vanishing field strength is,
\be
W_{\mu\nu}^3 = -E (\partial_\mu t \, \partial_\nu z - \partial_\nu t \, \partial_\mu z )
\ee
and $W_{\mu\nu}^1=0=W_{\mu\nu}^2$.

The currents are found from the classical equations of motion for the gauge fields,
\be
j^{\mu a} = \partial_\nu W^{\mu\nu a} + g\epsilon^{abc} W_\nu^b W^{\mu\nu c}, 
\label{jeq}
\ee
and (in the BW gauge) are given by
\be
j^{\mu 1} = - g \frac{E^2}{\Omega} \partial^\mu t , \ \
j^{\mu 2} = -  \Omega E \, \partial^\mu z , \ \
j^{\mu 3} =  0 .
\label{currents}
\ee

\subsection{Temporal gauge}

To bring the BW gauge fields to temporal gauge we perform the gauge
transformation, 
\be
U = e^{ i \sigma^2 \pi/4} e^{i\sigma^1 \pi/4}  e^{-i\sigma^1 \Omega t/2} 
\ee
where $\sigma^a$ are the Pauli spin matrices. Then, 
\be
W_\mu \to W_\mu'= U W_\mu U^\dag + \frac{2i}{g} U \partial_\mu U^\dag
\ee
where $W_\mu \equiv W_\mu^a \sigma^a$. This gives the gauge fields for a homogeneous 
electric field in the form,
\be
W^\pm_\mu = -\frac{\epsilon}{g}  e^{\pm i \Omega t} \, \partial_\mu z , \ \ 
W^3_\mu =0, 
\label{hombkgndpm}
\ee
where
\be
\epsilon \equiv g E/\Omega .
\label{epsE}
\ee
Next we also include a string profile function, $f(r)$, since we eventually
want to discuss electric string configurations.
Then the gauge fields we will consider are
\be
W^\pm_\mu = -\frac{\epsilon}{g}  e^{\pm i \Omega t} f(r) \, \partial_\mu z , \ \ 
W^3_\mu =0, 
\label{bkgndpm}
\ee
where $W^\pm_\mu \equiv W^1_\mu \pm i W^2_\mu$, $r$ refers to the
cylindrical radial coordinate.
Alternately we can write,
\ba
W^1_\mu &=& -\frac{\epsilon}{g}  \cos( \Omega t ) f(r) \, \partial_\mu z ,
\label{W1} \\ 
W^2_\mu &=& -\frac{\epsilon}{g}  \sin( \Omega t ) f(r) \, \partial_\mu z .
\label{W2}
\ea


The field strengths are
\ba
W^1_{\mu\nu} &=& - \frac{\epsilon}{g} \bigl [ 
-\Omega \sin(\Omega t) f(r) (\partial_\mu t \, \partial_\nu z - \partial_\nu t \, \partial_\mu z ) 
\nn \\
&& \hskip 0.5 cm
+ \cos(\Omega t) f'(r) (\partial_\mu r \, \partial_\nu z - \partial_\nu r \, \partial_\mu z ) \bigr ] 
\label{Wmunu1} \\
W^2_{\mu\nu} &=& - \frac{\epsilon}{g} \bigl [ 
\Omega \cos(\Omega t) f(r) (\partial_\mu t \, \partial_\nu z - \partial_\nu t \, \partial_\mu z ) 
\nn \\
&& \hskip 0.5 cm
+ \sin(\Omega t) f'(r) (\partial_\mu r \, \partial_\nu z - \partial_\nu r \, \partial_\mu z ) \bigr ] 
\label{Wmunu2} \\
W^3_{\mu\nu} &=& 0 
\label{Wmunu3}
\ea
where prime denotes differentiation with respect to the argument. Note that the electric field
is accompanied by a magnetic field in the azimuthal direction. To decide if the field configuration
is electric or magnetic we will calculate the gauge and Lorentz invariant Lagrangian density, 
\be
\Lag_g = - \frac{1}{4} W_{\mu\nu}^aW^{\mu\nu a}
= \frac{\epsilon^2}{2g^2} ( \Omega^2 f^2 - f'^2 )
\label{LW}
\ee
Positive $\Lag_g$ implies that the field is electric-like while negative values imply a
magnetic-like field.

The $SU(2)$ currents are obtained from \eqref{jeq},
\ba
j_\mu^1 &=& - \frac{\epsilon}{g} \cos(\Omega t) \left [ f'' + \frac{f'}{r} + \Omega^2 f \right ] \partial_\mu z
\label{j1} \\
j_\mu^2 &=& - \frac{\epsilon}{g} \sin(\Omega t) \left [ f'' + \frac{f'}{r} + \Omega^2 f \right ] \partial_\mu z 
\label{j2} \\
j_\mu^3 &=& - \frac{\epsilon^2}{g} \Omega f^2 \partial_\mu t 
\label{j3}
\ea
Note that the $a=1,2$ equations may also be written as,
\be
j_\mu^a =  \frac{1}{f} \left [ f'' + \frac{f'}{r} + \Omega^2 f \right ] W_\mu^a,  \ \ (a=1,2).
\label{ja12}
\ee

We are interested in finding if these currents can be sourced by scalar fields. We will
first consider the simpler case of a homogeneous electric field ($f(r)=1$) and show
that it
can be sourced by a scalar field in the fundamental representation. The case of a scalar
field in the adjoint representation is discussed in Appendix~\ref{adjoint} with the conclusion
that it {\it cannot} source the electric field.

\section{Homogeneous electric field}
\label{homoE}

In this section $f(r)=1$.

\subsection{Fundamental scalar}
\label{fundamental}

The model now is similar to the electroweak model where the symmetry is $SU(2)\times U(1)$,
except that the $U(1)$ charge, commonly denoted by $g'$, is set to zero. In other words, 
the $SU(2)$ is a gauged symmetry while the $U(1)$ is a global symmetry. 

We denote the fundamental scalar field by $\Phi$. The Lagrangian for the model is,
\be
L = - \frac{1}{4}W_{\mu\nu}^a W^{\mu\nu a} + | D_\mu\Phi |^2 - V(\Phi)
\label{Lag}
\ee
where
\be
D_\mu \Phi = \partial_\mu \Phi - i \frac{g}{2} W_\mu^a \sigma^a \Phi
\ee
\be
V(\Phi ) = m^2 |\Phi |^2 + \lambda |\Phi |^4
\label{vphi}
\ee
where $m^2$ may be negative or positive but $\lambda > 0$.

The gauge field equations of motion are
\be
D_\nu W^{\mu\nu a} = j_\mu^a = i \frac{g}{2} \left [ \Phi^\dag \sigma^a D_\mu \Phi - h.c. \right ],
\ee
where $h.c.$ stands for Hermitian conjugate.

Using $\sigma^a\sigma^b = \delta^{ab}+ i \epsilon^{abc}\sigma^c$ we find
\ba
&& \hskip -1 cm
\Phi^\dag \sigma^a D_\mu\Phi = \nn \\
&&
\Phi^\dag \sigma^a \partial_\mu \Phi -\frac{ig}{2}|\Phi|^2 W_\mu^a 
+ \frac{g}{2} |\Phi|^2 ({\vec W}_\mu\times {\hat n})^a
\label{phisigmadphi}
\ea
where the unit vector $\hatna$ is given by
\be
\hatna \equiv \frac{\Phi^\dag \sigma^a \Phi}{\Phi^\dag \Phi}.
\ee
Inserting \eqref{phisigmadphi} in the expression for the current, we get
\be
j_\mu^a = i \frac{g}{2} \left [ \Phi^\dag \sigma^a \partial_\mu \Phi - h.c. \right ] 
+ \frac{g^2}{2} |\Phi |^2 W_\mu^a .
\label{jaPhi}
\ee

We will start by solving \eqref{jaPhi} for $\Phi$ with $j_\mu^a$ given in 
\eqref{j1}-\eqref{j3} with $f=1$. We can write $\Phi$ in the Hopf parametrization,
\be
\Phi = \eta \begin{pmatrix} \cos\alpha \, e^{i\beta} \\ \sin\alpha \, e^{i\gamma} \end{pmatrix}.
\ee
We will assume that $\Phi$ is homogeneous, so $\alpha$, $\beta$ and $\gamma$
are only functions of time, and $\eta$ is a constant.

The $\mu =3$ component of \eqref{jaPhi} is non-trivial only for $a=1,2$, giving
\be
j_z^a = \frac{g^2}{2} |\Phi |^2 W_z^a , \ \ (a=1,2).
\label{jaPhiz}
\ee
Comparison with \eqref{ja12} gives,
\be
\Omega^2 = \frac{1}{2} g^2\eta^2 , \ \ {\rm or}, \ \ \eta = \frac{\sqrt{2}\, \Omega}{g}
\label{Omsoln}
\ee
(Recall that we are considering $f=1$.)
Next we turn to the $\mu=0$ components. Some algebra (see Appendix~\ref{someformulae})
leads to
\ba
2{\dot \alpha} \sin(\gamma-\beta) + \sin(2\alpha) \cos(\gamma-\beta) ({\dot \beta}+{\dot \gamma}) &=&0\\
2{\dot \alpha} \cos(\gamma-\beta) - \sin(2\alpha) \sin(\gamma-\beta) ({\dot \beta}+{\dot \gamma}) &=&0\\
g\eta^2 (\cos^2\alpha \, {\dot \beta} - \sin^2\alpha \, {\dot \gamma} ) - \frac{\epsilon^2 \Omega}{g} &=& 0
\ea

The solution to these three equations leads to
\be
\Phi = \eta \begin{pmatrix} z_1 e^{+i\omega t} \\ z_2 e^{-i\omega t} \end{pmatrix}
\label{Phisolgp0}
\ee
where $z_1,\, z_2 \in \mathbb{C}$ are constants with $|z_1|^2+|z_2|^2=1$, and
\be
\omega = \frac{\epsilon^2}{2\Omega}
\label{omsol}
\ee

Now we have to make sure that $\Phi$ solves its own equation of motion,
\be
D_\mu D^\mu \Phi + V'(\Phi) =0
\label{Phieom}
\ee
where the prime denotes derivative with respect to $\Phi^\dag$. 
%
%
For $V$ in \eqref{vphi} and $\Phi$ in \eqref{Phisolgp0}, we can write
\be
V'(\Phi ) = (m^2 + 2 \lambda \eta^2 ) \Phi.
\label{V'Phi}
\ee
We also evaluate
\be
D_\mu D^\mu \Phi  = - \left ( \omega^2 - \frac{\epsilon^2}{4} \right ) \Phi
\ee
Therefore \eqref{Phieom} leads to the equation,
\be
-\omega^2 + \frac{\epsilon^2}{4} + m^2 + 2 \lambda \eta^2  = 0.
\ee
Consistency with \eqref{omsol} implies
\be
\omega^2 - \frac{\omega \Omega}{2} - m^2 - 2 \lambda \eta^2 = 0.
\ee
Therefore
\be
\omega = \frac{1}{2} \left [ \frac{\Omega}{2} \pm 
\sqrt{\frac{\Omega^2}{4} + 4 \left (m^2 + \frac{4\lambda}{g^2} \Omega^2 \right )} \, \right ] .
\label{omsol2}
\ee
Note that \eqref{omsol} implies that $\omega/\Omega > 0$. Depending on the signs and
magnitude of the parameters $m^2$, $\lambda/g^2$ and $\Omega^2$,
only one or both roots in \eqref{omsol2} are valid.

If we choose potential parameters such that $m^2+2\lambda\eta^2=0$, we obtain
simpler expressions. With this choice of parameters Eq.~\eqref{V'Phi} gives $V'(\Phi )=0$, 
{\it i.e.} $\Phi$ is at an extrema of its potential.
Then \eqref{omsol2} gives two solutions: $\omega=0$ or $\omega = \Omega/2$.
The solution with $\omega=0$ is trivial for then $\epsilon =0$ because of \eqref{omsol} and
the electric field vanishes.
For the non-trivial solution $\omega=\Omega/2$,
\eqref{omsol} gives
\be
\epsilon = \Omega, \ \ (V'=0).
\ee

To summarize these results: we have found a solution of the classical equations
of motion
that corresponds to a homogeneous electric field,
\ba
\Phi &=& \frac{\sqrt{2}\, \Omega}{g} \begin{pmatrix} z_1 e^{+i\omega t} \\ z_2 e^{-i\omega t} \end{pmatrix}
\\
{\vec W}_\mu &=& -\frac{\sqrt{2\omega\Omega}}{g}  
\left (\cos( \Omega t ),\sin( \Omega t ),0 \right )\partial_\mu z ,
\label{summaryPhiW}
\ea
where $z_1,\, z_2 \in \mathbb{C}$ are constants with $|z_1|^2+|z_2|^2=1$ and $\omega$
is given in terms of $\Omega$ and the parameters in the scalar potential by \eqref{omsol2}.
The electric field is found from \eqref{epsE} and \eqref{omsol},
\be
E = \frac{\Omega \sqrt{2\omega \Omega}}{g}
\label{Esol}
\ee

\section{Electric string ($f(r) \ne 1$)}
\label{estring}

We now move on from the homogeneous electric field to electric string solutions.
The gauge fields are given by \eqref{bkgndpm}, the required currents by \eqref{j1}-\eqref{j3},
and the currents that $\Phi$ can source by \eqref{jaPhi}.
%
%
Hence we must now find $\Phi$ by solving,
\ba
&& \hskip -0.5 cm
\frac{i}{2} g ( \Phi^\dag \sigma^1 D_\mu \Phi - (D_\mu \Phi)^\dag \sigma^1 \Phi ) \nn \\
&& \hskip 1 cm
=  -\frac{\epsilon}{g} \cos(\Omega t) \left ( f'' + \frac{f'}{r} + \Omega^2 f \right ) \partial_\mu z 
\label{j1eq} \\
&&\hskip -0.5 cm
\frac{i}{2} g ( \Phi^\dag \sigma^2 D_\mu \Phi - (D_\mu \Phi)^\dag \sigma^2 \Phi ) \nn \\
&& \hskip 1 cm
=  -\frac{\epsilon}{g} \sin(\Omega t) \left ( f'' + \frac{f'}{r} + \Omega^2 f \right ) \partial_\mu z 
\label{j2eq} \\
&&\hskip -0.5 cm
\frac{i}{2} g ( \Phi^\dag \sigma^3 D_\mu \Phi - (D_\mu \Phi)^\dag \sigma^3 \Phi ) 
= -\frac{\epsilon^2}{g} \Omega  f^2 \, \partial_\mu t 
\label{j3eq} 
\ea

Guided by the homogeneous electric field case of Sec.~\ref{homoE} we try,
\be
\Phi = \eta h(r) \begin{pmatrix} z_1 e^{+i\omega t} \\ z_2 e^{-i\omega t} \end{pmatrix}
\label{Phiansatz1}
\ee
where $h(r)$ is a real profile function that is to be determined.

We first consider $\mu=0$ in \eqref{j1eq}-\eqref{j3eq}.
We find that \eqref{j1eq} and \eqref{j2eq} are trivially satisfied 
while Eq.~\eqref{j3eq} gives
\be
-g\eta^2 h^2 \omega = - \frac{\epsilon^2}{g} \Omega f^2
\label{relation1}
\ee
which implies
\be
h(r) = \frac{\epsilon}{g\eta} \sqrt{\frac{\Omega}{\omega}} \, f(r)
\label{hf}
\ee
and we should have $\omega/\Omega > 0$ since $h$ is real.

The $\mu=1,2$ equations are trivially satisfied, so we now
consider $\mu=3$, {\it i.e.} the $\mu=z$ equations. Eqs.~\eqref{j1eq} and \eqref{j2eq}
then give
\be
 f'' + \frac{f'}{r} + \left ( \Omega^2 - \frac{g^2  \eta^2}{2} h^2 \right ) f = 0
 \label{hfdiffeq}
\ee
and \eqref{j3eq} is trivially satisfied.

Inserting \eqref{hf} in \eqref{hfdiffeq} gives
\be
 f'' + \frac{f'}{r} + \Omega^2 \left ( 1 - \frac{\epsilon^2 }{2\omega  \Omega} f^2 \right ) f = 0.
 \label{hfdiffeq2}
\ee

Next we consider the $\Phi$ equation of motion in \eqref{Phieom} together
with \eqref{hf} to get
\be
 f'' + \frac{f'}{r} +  \left [ (\omega^2 -m^2)  - \left (  \frac{\epsilon^2}{4} 
 + \frac{2\lambda \epsilon^2 \Omega}{g^2 \omega} \right ) f^2 \right ] f = 0.
 \label{alsofeq}
\ee
Consistency with \eqref{hfdiffeq} requires
\ba
\Omega^2 &=& \omega^2 - m^2
\label{match1} \\
\frac{\Omega}{2 \omega} &=&  \frac{1}{4} + \frac{2\lambda \Omega}{g^2 \omega} .
\label{match2}
\ea

Eq.~\eqref{match2} can be written as
\be
\omega = 2 \left ( 1- \frac{4\lambda}{g^2} \right ) \Omega
\label{omOmlambda}
\ee
and since we have already assumed $\omega/\Omega > 0$ (see \eqref{hf})
we should restrict to $\lambda \le g^2/4$. Also since the potential $V$ should be
bounded from below, we have $0 \le \lambda \le g^2/4$.

For fixed parameters $m^2$, $\lambda$, Eqs.~\eqref{match1} and \eqref{match2} can be
solved to obtain $\Omega$ and $\omega$,
\ba
\Omega^2 = \frac{m^2}{4(1-4\lambda/g^2)^2-1} \label{Om2sol} \\
\omega^2 =  \frac{4(1-4\lambda/g^2)^2 m^2}{4(1-4\lambda/g^2)^2-1} \label{om2sol}
\ea
This is a valid solution provided
\be
 \frac{m^2}{4(1-4\lambda/g^2)^2-1} > 0
\ee
which gives the conditions $0 < \lambda < g^2/8$ or $\lambda > 3g^2/8$ if $m^2 > 0$, 
and $g^2/8 < \lambda < 3g^2/8$ 
if $m^2 <0$. Taking into account the tighter restriction discussed below \eqref{omOmlambda}
that $\omega >0$ for $\Omega > 0$, the 
range of $\lambda$ for $m^2 < 0$ is $g^2/8 < \lambda < g^2/4$.

\begin{figure}
\includegraphics[width=0.5\textwidth,angle=0]{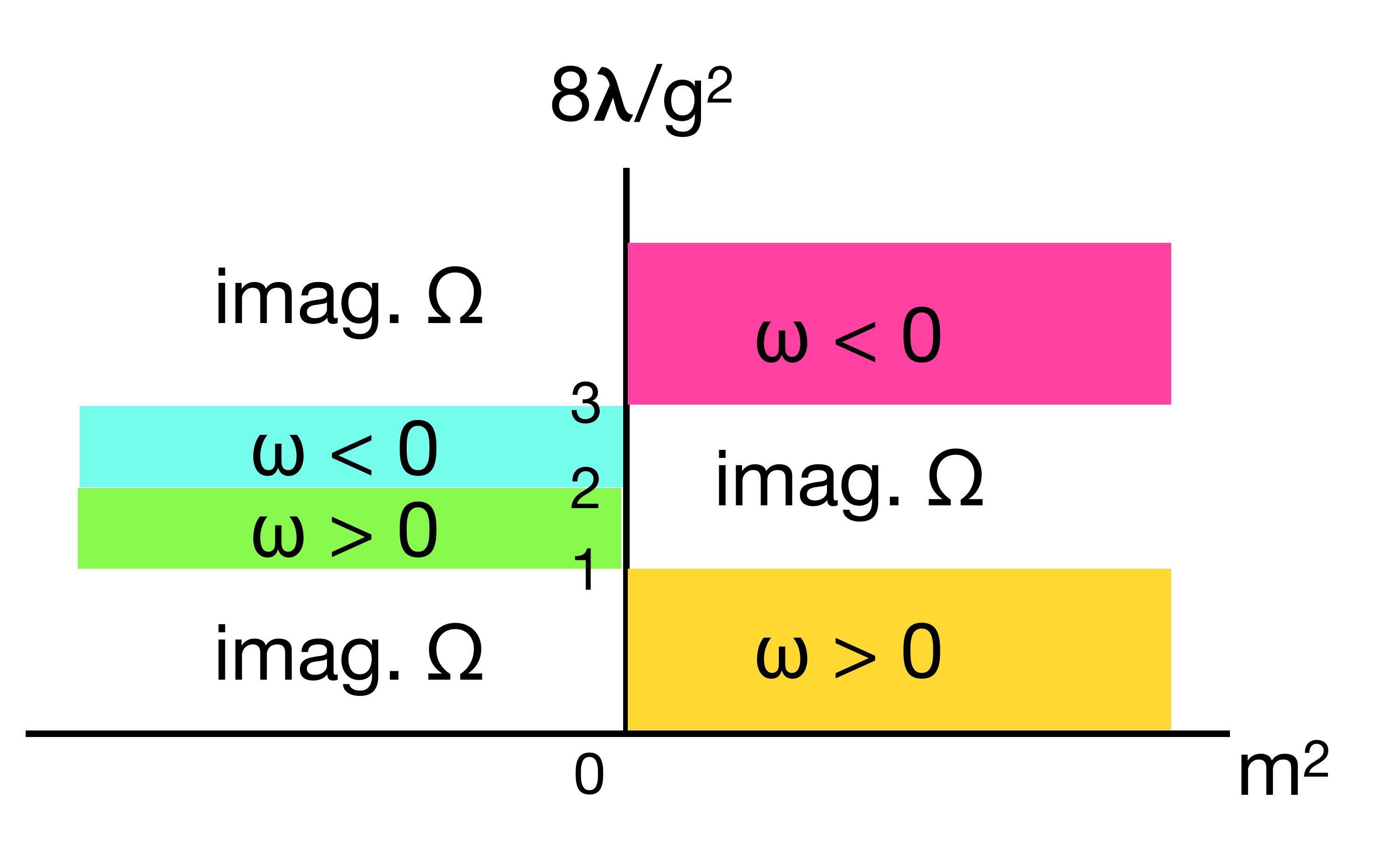}
 \caption{Constraints on the parameters in the $m^2$-$8\lambda/g^2$ plane. The unshaded
 regions give imaginary $\Omega$ and are not allowed. 
 The solution is only valid in the regions of parameter space where $\omega > 0$.
 }
\label{constraints}
\end{figure}

Hence there is a range of values of the parameters $m^2$ and $\lambda$ of the potential for which
\eqref{hfdiffeq2} and \eqref{alsofeq} are identical, provided \eqref{Om2sol} and \eqref{om2sol} 
hold.
In this case, \eqref{hfdiffeq2} can be solved numerically. It is 
preferable to rescale variables,
\be
R = \Omega r, \ \ F = \frac{\epsilon}{\sqrt{2\omega \Omega}} f
\ee
and then
\be
F'' + \frac{F'}{R} +  (1 -  F^2 ) F = 0.
 \label{Fdiffeq}
\ee
Only the combination $\epsilon f(r)$ appears in the gauge field (see \eqref{bkgndpm}), 
and we are free to take $f(0)=1$ or,
\be
F(0)= 
\frac{\epsilon}{\sqrt{2\omega \Omega}} .
\ee
Since the electric field strength is proportional to $F$, we would like to choose boundary
conditions such that $F(0) = F_0 \ne 0$ and $F(\infty ) \to 0$. Different values of $F_0$
correspond to different external charges placed at $z \to \pm \infty$ that produce the
electric field. Smoothness at $R=0$ requires $F'(0)=0$. 

\begin{figure}
\includegraphics[width=0.4\textwidth,angle=0]{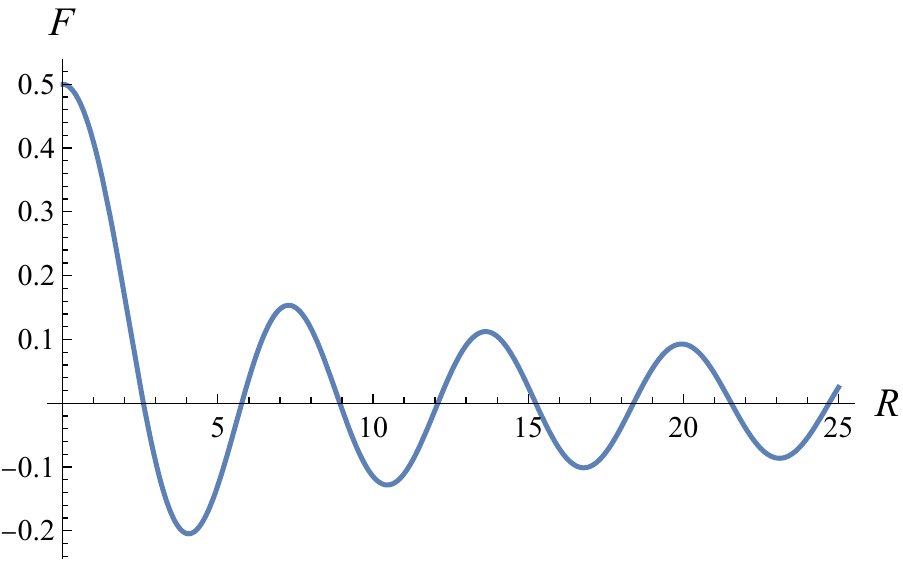}
 \caption{$F(R)$ vs. $R$ for $F_0=0.5$.}
\label{FR}
\end{figure}

There are no solutions that fall-off asymptotically for $F_0 > 1$. For
$F_0 \ll 1$, the non-linear term with $F^2$ is sub-dominant and the solutions are well
approximated by a Bessel function of zero order,
\be
F(R) \approx F_0 J_0 (R).
\ee
A plot of the numerically evaluated $F(R)$ is shown in Fig.~\ref{FR}.
The asymptotic behavior of $F(R)$ is therefore,
\be
F(R) \sim F_0 \sqrt{\frac{2}{\pi R}} \cos \left ( R -\frac{\pi}{4} \right ).
\label{asympF}
\ee

\begin{figure}
\includegraphics[width=0.4\textwidth,angle=0]{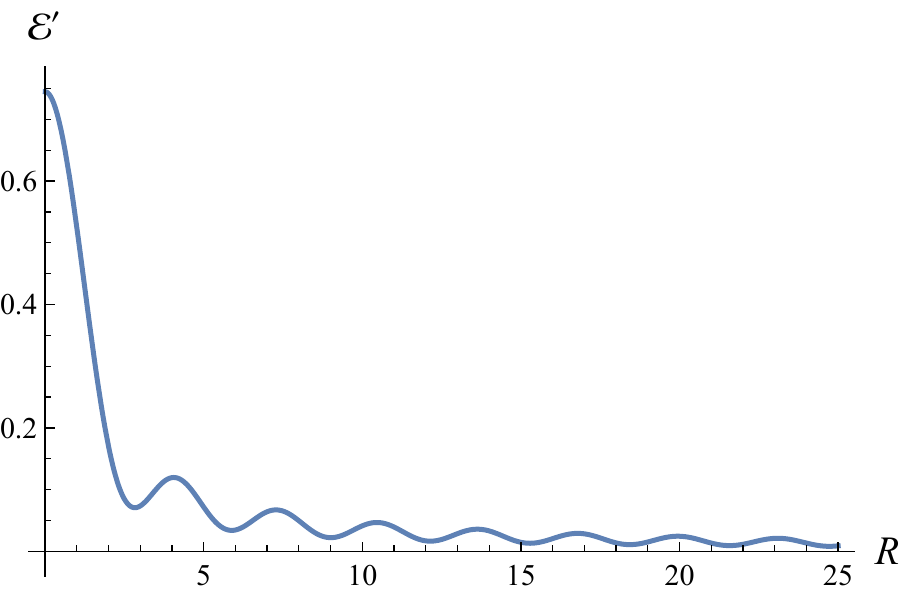}
 \caption{$\calE'$ vs. $R$ for $F_0=0.5$ and $\kappa =1.5$.}
\label{calEplot}
\end{figure}

The energy density in all the fields is given by the expression,
\be
\calE = \frac{1}{2} (W_{0i}^a )^2 +  \frac{1}{4} (W_{ij}^a )^2 + 
|D_t\Phi |^2 + |D_i\Phi |^2 + V(\Phi )
\ee
Inserting the expressions for the solution we get
\ba
\calE &=& 
\frac{\epsilon^2 \Omega^2 }{2 g^2} f^2 + \frac{\epsilon^2}{2 g^2}  f'^2
+ \frac{\epsilon^2 \omega \Omega}{g^2} f^2
+ \frac{\epsilon^2 \Omega}{g^2\omega} f'^2 
+ \frac{\epsilon^4\Omega}{4g^2\omega} f^4
 \nn \\
&&
+m^2 \frac{\epsilon^2 \Omega}{g^2 \omega} f^2
+\lambda \frac{\epsilon^4 \Omega^2}{g^4 \omega^2} f^4,
\ea
or in terms of rescaled variables,
\ba
\calE' \equiv \frac{g^2}{2\omega \Omega^3}\calE
&=& \left ( \frac{1}{2} + \frac{1}{\kappa} \right ) F'^2 +
\left ( \frac{1}{2} + 2\kappa - \frac{1}{\kappa} \right ) F^2 \nn \\
&&
+  \frac{1}{2} \left ( \frac{1}{2} + \frac{1}{\kappa} \right ) F^4
\ea
where (see \eqref{omOmlambda}),
\be
\kappa \equiv \frac{\omega}{\Omega} = 2 \left ( 1- \frac{4\lambda}{g^2} \right ).
\ee
Due to the constraints on $\lambda/g^2$, we find $1 < \kappa < 2$ for $m^2 > 0$
and $0 < \kappa < 1$ for $m^2 < 0$.
In Fig.~\ref{calEplot} we show $\calE'$ for a sample value of $\kappa$
and $F_0$.

The energy per unit length along the $z-$direction is defined to be the tension
of the string. Therefore the rescaled tension with a radial cutoff at $R_c$ is,
\ba
\mu(R_c) &=& 2\pi \int_0^{R_c} dR \, R \, \calE'  \nn \\
&\approx& \left [ \left ( \frac{1}{2} + \frac{1}{\kappa} \right ) 0.54 +
\left ( \frac{1}{2} + 2\kappa - \frac{1}{\kappa} \right ) 0.54 \right ] R_c \nn \\
&&
+  \frac{1}{2} \left ( \frac{1}{2} + \frac{1}{\kappa} \right ) 0.09\, \ln (R_c).
\ea
Therefore the tension diverges linearly with the radial cutoff.
Thus while the electric field is axisymmetric and concentrated along the
central axis, it is not as sharply localized as in a magnetic Nielsen-Olesen
string~\cite{Nielsen:1973cs}, though it is more localized than the homogeneous
configuration. The situation is similar to that of a global string for which
the energy diverges logarithmically and to global monopoles for which
the energy diverges linearly~\cite{Vilenkin:2000jqa}.

For $m^2 < 0$, the potential $V(\Phi)$ has a minimum at $\Phi \ne 0$.
However the solution of \eqref{Fdiffeq} implies that $\Phi \to 0$ as
$r \to \infty$. Therefore
$\Phi$ is not in its true vacuum asymptotically and there is non-vanishing
vacuum energy at spatial infinity. 
In this case the electric string is a cylindrical ``bubble'' of the true
vacuum (with non-vanishing $\Phi$) in a background of the false 
vacuum  phase with $\Phi=0$.
This is different from the case of the electric string with $m^2 > 0$ for 
then the true vacuum is at $\Phi =0$ and the
potential energy of $\Phi$ goes to zero in the asymptotic region.

\begin{figure}
\includegraphics[width=0.4\textwidth,angle=0]{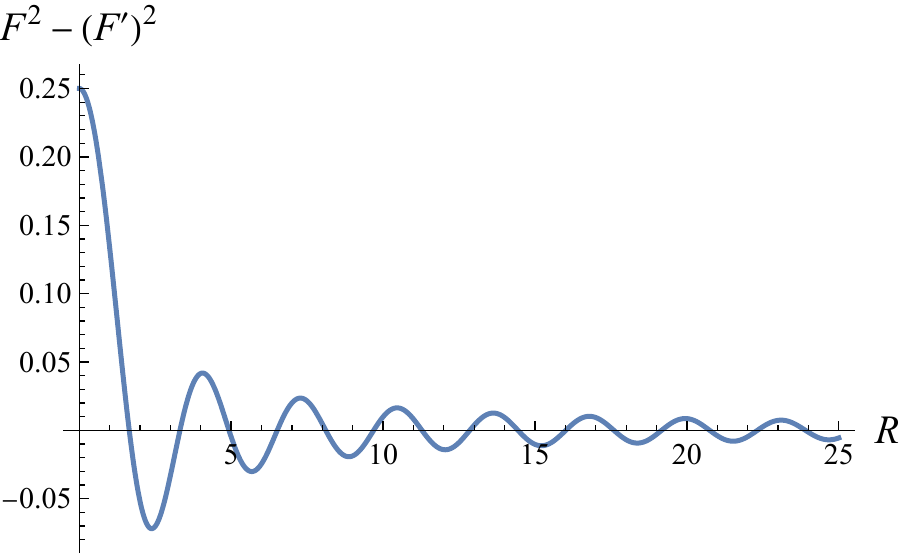}
 \caption{$F(R)^2-F'(R)^2$ vs. $R$ for $F_0=0.5$. The field strength is
electric-like where $F(R)^2-F'(R)^2$ is positive and magnetic-like where
 $F(R)^2-F'(R)^2$ is negative.}
\label{FLag}
\end{figure}

\begin{figure}
\includegraphics[width=0.4\textwidth,angle=0]{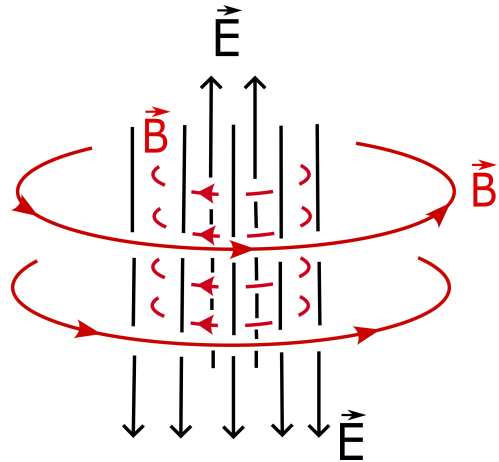}
 \caption{A sketch of the electric and magnetic fields in an electric string.}
\label{sketch}
\end{figure}

Finally we return to the question of whether the solution corresponds to
an {\it electric} string as azimuthal magnetic fields are also present
(see \eqref{Wmunu1}-\eqref{Wmunu3}). Hence we calculate $\Lag_g$ 
using \eqref{LW} and plot the quantity $F^2-(F')^2$ in Fig.~\ref{FLag}.
The behavior at large $R$ can be seen from the properties of the Bessel 
functions,
\be
F^2-(F')^2 \sim F_0^2 [ J_0^2(R) - J_1^2(R) ] \to F_0^2 \frac{2\sin(2R)}{\pi R}
\ee
where we have used $J_1(R) = -J_0'(R)$ and \eqref{asympF}. 
The gauge field strength is electric-like where $F^2-(F')^2 > 0$, otherwise
it is magnetic-like. This shows that the solution has alternating electric
and magnetic fields where the electric field is along the $z-$ direction
and the magnetic field is along the azimuthal direction.
So a caricature of the electric string configuration is a tube of electric
field along the $z$ direction, wrapped by weak azimuthal magnetic fields, 
that are again contained in a sheath of weaker electric field, ad infinitum
(see the sketch in Fig.~\ref{sketch}).

To summarize the main result of this section, we have found the electric string
solution,
\ba
\Phi &=& \frac{\epsilon}{g} \sqrt{\frac{\Omega}{\omega}} \, f(r)
\begin{pmatrix} z_1 e^{+i\omega t} \\ z_2 e^{-i\omega t} \end{pmatrix}
\label{stringPhi} \\
W^\pm_\mu &=& -\frac{\epsilon}{g}  e^{\pm i \Omega t} f(r) \, \partial_\mu z , \ \ 
W^3_\mu =0, 
\label{stringW}
\ea
where $|z_1|^2+|z_2|^2=1$, and $\Omega$ and $\omega$ are given by
\ba
\Omega = \left [ \frac{m^2}{4(1-4\lambda/g^2)^2-1} \right ]^{1/2} \label{Om2sol2} \\
\omega = \left [  \frac{4(1-4\lambda/g^2)^2 m^2}{4(1-4\lambda/g^2)^2-1} \right ]^{1/2} \label{om2sol2}
\ea
and $m^2$ and $\lambda$ are parameters of the scalar potential (see \eqref{vphi}).
The solution is valid for $0 < \lambda < g^2/8$ for $m^2 > 0$ and for $g^2/8 < \lambda < g^2/4$
for $m^2 < 0$. The profile function, $f(r)$, is common to both the gauge field and the scalar
field and satisfies \eqref{hfdiffeq2} with boundary conditions $f(0)=1$, $f'(0)=0$. The solution 
for the profile is closely approximated by the zeroth order Bessel function,
$J_0(\Omega r)$, up to a multiplicative constant that fixes the strength of the electric field
at the origin.

%
%
%

\section{Schwinger gluon production?}
\label{schwinger}

The gauge particles (``gluons'') are charged under $SU(2)$, as are excitations
of the scalar field $\Phi$, and there can be Schwinger pair production of both
kinds of excitations. Here we are interested in whether the electric field is 
protected from pair production of gluons. Pair production of scalar quanta
is similar to pair production of quarks in QCD that results in string breaking.
The effect can be suppressed by considering large masses of the scalar
quanta, {\it i.e.} large positive values of the parameter $m^2$.

To see that the electric field is stable to Schwinger gluon production, we
perturb the gauge field,
\be
W_\mu^\pm = A_\mu^\pm + e^{\pm i\Omega t} Q_\mu^\pm, \ \
W_\mu^3 = A_\mu^3 + Q_\mu^3
\label{WQ}
\ee
where $A_\mu^a$ is the background (see Eq.~\eqref{bkgndpm})
\be
A^\pm_\mu = -\frac{\epsilon}{g}  e^{\pm i \Omega t} f(r) \, \partial_\mu z , \ \ 
A^3_\mu =0
\label{bkgndpmA}
\ee
and $Q_\mu^a$ are the perturbations.
The scalar field is left unperturbed,
\be
\Phi = \eta h(r) \begin{pmatrix} z_1 e^{+i\omega t} \\ z_2 e^{-i\omega t} \end{pmatrix},
\ee
since we are interested in gluon production and the mass of the $\Phi$ field can be
taken to be large.

As in Ref.~\cite{Vachaspati:2022ktr}, the expressions in \eqref{WQ} are inserted
in the Lagrangian density for the gauge fields
\be
\Lag_g = - \frac{1}{4} W_{\mu\nu}^aW^{\mu\nu a}
\label{LagW}
\ee
to obtain the Lagrangian density for the perturbations $Q_\mu^a$. The expressions
are lengthy but the important point is that there is no explicit time-dependence in the
Lagrangian even though the background in \eqref{bkgndpmA} is time-dependent.
For example, the Lagrangian density to second order in the perturbations is
\ba
&& \hskip -0.75 cm 
\Lag_g^{(2)} = \nn \\
&& 
\half ( {\dot Q}^{(1)}_i - \Omega Q^{(2)}_i )^2
+\half ( {\dot Q}^{(2)}_i + \Omega Q^{(1)}_i )^2 
+\half ( {\dot Q}^{(3)}_i  )^2 \nn \\
&-& \fourth ( \partial_i Q^{(1)}_j -  \partial_j Q^{(1)}_i )^2 \nn \\
&-& \fourth ( \partial_i Q^{(2)}_j -  \partial_j Q^{(2)}_i + \epsilon f ({\hat z}_i Q^{(3)}_j -{\hat z}_j Q^{(3)}_i ))^2 \nn \\
&-& \fourth ( \partial_i Q^{(3)}_j -  \partial_j Q^{(3)}_i - \epsilon f ({\hat z}_i Q^{(2)}_j -{\hat z}_j Q^{(2)}_i ))^2 \nn \\
&+& \epsilon f' ({\hat z}_i {\hat r}_j - {\hat z}_j {\hat r}_i ) Q^{(2)}_i Q^{(3)}_j
\label{Lg2corrected}
\ea
where ${\hat z}_i$, ${\hat r}_i$ are unit vectors in the $z-$ and $r-$
directions, and the contraction of spatial indices is with the
Kronecker delta, {\it e.g.} $( {\dot Q}^{(3)}_i  )^2 =  {\dot Q}^{(3)}_i {\dot Q}^{(3)}_i $. 
Similar expressions are obtained at all orders in perturbations and there is no
explicit time-dependence in any of them.
If we expand the perturbations in modes in $\Lag_g^{(2)}$, the mode coefficients 
correspond to simple harmonic oscillators with time-independent frequencies, which
implies that there is no particle production.

In the present analysis, since we include the scalar field $\Phi$, there is an extra term in 
$Q_\mu^a$ coming from the covariant gradient term in the Lagrangian,
\be
L_\Phi = |D_\mu\Phi |^2 - V(\Phi ) \to \ldots + \frac{g^2}{4} |\Phi |^2 Q_\mu^a Q^{\mu a}
\label{LPhi}
\ee
where the $\ldots$ include terms that are zeroth and first order in $Q_\mu^a$.
Since $|\Phi |^2 = \eta^2 h^2$ is independent of time, the last term in \eqref{LPhi}
is simply a mass term for $Q_\mu^a$ with a mass that is independent of time.
Once again, the quantum state of the modes of $Q_\mu^a$ will correspond
to simple harmonic oscillators with time-independent frequencies. Thus the 
time dependence of the background gauge field does not lead to any Schwinger
particle production of the gauge excitations.

\section{Conclusions}
\label{conclusions}

We have first constructed homogeneous electric field solutions in non-Abelian
gauge theories with a scalar field that transforms in the fundamental representation.
This construction paved the way for the construction of {\it electric string} solutions
that are summarized in Eqs.~\eqref{stringPhi}, \eqref{stringW}, \eqref{Om2sol2}
and \eqref{om2sol2}. The solutions describe a flux tube of electric field, wrapped
by azimuthal magnetic fields, followed by a sheath of electric field, which is
again wrapped by azimuthal magnetic field, ad infinitum. The strength of the
electric and magnetic fields falls off with distance as $1/\sqrt{r}$.
The slow fall off implies a linear divergence in the energy per unit length of
the electric string for $m^2 >0$. In the case where $m^2 < 0$, the electric
solution has $\Phi=0$ asymptotically while the true vacuum has $\Phi \ne 0$.
Hence the electric string for $m^2 < 0$ is like a cylindrical bubble solution that 
contains gauge fields and $\Phi \ne 0$, that is immersed in a false vacuum region 
with $\Phi=0$.

In Sec.~\ref{schwinger} we have shown that Schwinger gluon
production is absent in the electric string background. This still leaves room 
for classical instabilities of the electric string solution, especially since
unstable modes are known to exist in the {\it homogeneous} BW gauge 
field background in pure gauge theory~ \cite{Pereira:2022lbl}. With the $\Phi$ field
included, the main difference is that there is now an interaction term
$|\Phi |^2 (W^a_i)^2$ in the energy functional. Since $\Phi$ is non-zero
in the electric string solution, the gauge field excitations above the
background are massive. This should suppress instabilities but it is difficult 
to say if the suppression is sufficient to eliminate the instabilities. We plan to 
perform a classical stability analysis of the electric string solution in future work.

There is a large body of work on the quantization of classical 
solutions~\cite{Rajaraman:1982is}
The procedure is to consider fluctuations around the background solution. In our
case, the gauge field with fluctuations can be written as in Eq.~\eqref{WQ} and similarly the
scalar field is,
\be
\Phi = \Phi_0 + \hat\Phi
\ee
where $\Phi_0$ is the classical solution and $\hat\Phi$ represent fluctuations. 
Assuming weak coupling and that there are no classical instabilities, the
fluctuations can be treated to lowest quadratic order in the action and their
eigenmodes are simple harmonic oscillators that can be quantized in the standard 
way. The backreaction of these quantum fluctuations on the classical background
will be small. However, this straightforward quantization does not hold at strong
coupling. In that case, the action cannot be truncated to quadratic order in
the fluctuations and the backreaction may change the classical solution in a
significant way. Then lattice methods seem to be the only recourse. It would 
be very interesting if strong coupling effects could control the asymptotic
behavior of the electric string so as to give a finite string tension.

\acknowledgements
I thank Jude Pereira for comments and for the drawing in Fig.~\ref{sketch}.
This work was supported by the U.S. Department of Energy, Office of High Energy 
Physics, under Award No.~DE-SC0019470.


\appendix

\section{Homogeneous electric field and adjoint scalar field}
\label{adjoint}

The $SU(2)$ gauge fields (see \eqref{W1} and \eqref{W2}) will be written as
\be
{\vec W}_\mu = -\frac{\epsilon}{g} (\cos(\Omega t), \sin(\Omega t), 0) \, \partial_\mu z
\label{Wso3}
\ee
where the vector sign denotes a vector in internal space.
From \eqref{j1}-\eqref{j3}, the current is
\be
{\vec j}_\mu = -\frac{\epsilon \Omega}{g} ( \Omega\cos(\Omega t) \partial_\mu z, 
\Omega \sin(\Omega t)\partial_\mu z, \epsilon \partial_\mu t )
\label{wantedj}
\ee

The adjoint scalar will be denoted by ${\vec \phi}$.
In terms of ${\vec \phi}$ the current is
\be
{\vec j}_\mu = g\, {\vec \phi} \times D_\mu {\vec \phi}
\label{jphiDphi}
\ee
where
\be
D_\mu {\vec \phi} = \partial_\mu {\vec \phi} + g {\vec W}_\mu \times {\vec \phi}
\ee
Therefore, given ${\vec j}_\mu$, ${\vec \phi}$ must satisfy the constraint
\be
{\vec \phi}\cdot {\vec j}_\mu =0
\ee
for every $\mu$. Setting $\mu=0$ we obtain the requirement $\phi^3 =0$.
And setting $\mu = 3$ gives
\be
\cos(\Omega t) \phi^1 + \sin(\Omega t) \phi^2 =0
\ee
Therefore
\be
{\vec \phi} = \eta (-\sin(\Omega t), \cos (\Omega t), 0)
\ee
where $\eta$ is some unspecified vacuum expectation value of the scalar.
Then
\be
\partial_\mu {\vec \phi} = - \eta \Omega \, (\cos(\Omega t), \sin (\Omega t), 0) \partial_\mu t
\ee
and
\be
{\vec \phi} \times \partial_\mu {\vec \phi} = \eta^2 \Omega \, (0,0,1) \partial_\mu t
\ee
\be
{\vec \phi}\times ({\vec W}_\mu \times {\vec \phi})= 
({\vec \phi}\cdot {\vec \phi}) {\vec W}_\mu - ({\vec \phi}\cdot {\vec W}_\mu ) {\vec \phi} =
\eta^2 {\vec W}_\mu,
\ee
since ${\vec \phi}\cdot {\vec W}_\mu =0$.

Eq.~\eqref{jphiDphi} gives
\ba
{\vec j}_\mu &=& g\eta^2 \left [ \Omega {\hat e}_3 \partial_\mu t + g {\vec W}_\mu \right ] \nn \\
&& \hskip -0.75 cm
= g\eta^2 (-\epsilon \cos(\Omega t) \partial_\mu z, - \epsilon \sin(\Omega t) \partial_\mu z,
\Omega \partial_\mu t)
\label{j3sol}
\ea
Comparison of the $\mu=3$ expressions with the desired currents \eqref{wantedj} gives,
\be
\Omega = g\eta
\ee
The trouble arises in matching the $\mu =0$ expressions, for then,
\be
\epsilon^2 = - g^2\eta^2
\ee
and a real solution does not exist.

We conclude that the adjoint scalar ${\vec \phi}$ cannot source the initial gauge field
in \eqref{Wso3}.

\section{Homogeneous electric field and electroweak model}
\label{electroweak}

The electroweak model has the same ingredients as our model with an electric
string solution, except that the $U(1)$ symmetry is gauged with gauge coupling
$g'$ and we have an extra gauge field, $Y_\mu$, called the hypercharge gauge field. 
We will look for an electric string solution of the same form as in \eqref{bkgndpm} and 
with $Y_\mu =0$.

The $W$ currents are unchanged and we still need to satisfy \eqref{j1eq}-\eqref{j3eq}.
In addition, since $Y_\mu=0$, the hypercharge current must also vanish,
\be
\partial_\nu Y^{\mu\nu}  = j_\mu^Y = i \frac{g'}{2} \left ( \Phi^\dag D_\mu \Phi - h.c. \right ) =0
\label{jY}
\ee
The form of $\Phi$ and $W_\mu^a$ is fixed and given in \eqref{stringPhi} and \eqref{stringW}.
Inserting these in \eqref{jY} gives,
\ba
0 &=& -\frac{g' \epsilon^2}{g^2} \Omega f^2 (|z_1|^2-|z_2|^2) \partial_\mu t 
\nn \\
&& 
-\frac{g' \epsilon^3}{2g^2} \frac{\Omega}{\omega} f^3 
\left [ z_1 z_2^* e^{i(\Omega +2\omega)t} + c.c. \right ] \partial_\mu z,
\label{jmuY}
\ea
The $\mu=0$ component requires $|z_1|^2=|z_2|^2 = 1/2$, while the
$\mu = z$ component requires,
\be
z_1 z_2^* e^{i(\Omega +2\omega)t} + c.c. =0 .
\ee
This forces $\omega/\Omega = -1/2$ but this is in conflict with the requirement
that $\omega /\Omega > 0$ discussed below \eqref{hf}. Hence the electric string
solution with $Y_\mu=0$ does not exist in the electroweak model. This does not
exclude the possibility of an electric string solution with $Y_\mu \ne 0$.

\section{Some helpful formulae}
\label{someformulae}

If we use the Hopf parametrization to write $\Phi$,
\be
\Phi = \eta \begin{pmatrix} \cos\alpha \,  e^{i\beta}\\ \sin\alpha \, e^{i\gamma} \end{pmatrix}
\ee
where $\alpha$, $\beta$, $\gamma$ only depend on time.
Then
\ba
&& \hskip -1.25 cm
{\vec n} \equiv \Phi^\dag {\vec \sigma} \Phi
= \eta^2 (\sin(2\alpha)\cos\theta, \sin(2\alpha) \sin\theta,\cos(2\alpha))
\ea
where $\theta \equiv \gamma-\beta$. And
\ba
&& \hskip -1 cm
\Phi^\dag \sigma^1 {\dot \Phi} - h.c. = i\eta^2 [ 2\sin\theta \, {\dot \alpha} 
+ \sin2\alpha \cos\theta ({\dot \beta}+{\dot \gamma}) ] 
\\ &&  \hskip -1 cm
\Phi^\dag \sigma^2 {\dot \Phi} - h.c. = i\eta^2 [ -2\cos\theta\, {\dot \alpha} 
+ \sin2\alpha \sin\theta ({\dot \beta}+{\dot \gamma}) ]
\\ &&  \hskip -1 cm
\Phi^\dag \sigma^3 {\dot \Phi} - h.c. = i\eta^2 2 (\cos^2\alpha \, {\dot \beta}-\sin^2\alpha \, {\dot \gamma} )
\\ &&  \hskip -1 cm
\Phi^\dag  {\dot \Phi} - h.c. = i\eta^2 2 (\cos^2\alpha \, {\dot \beta}+ \sin^2\alpha \, {\dot \gamma} )
\ea


\bibstyle{aps}
\bibliography{paper}

\end{document}